\documentstyle[12pt]{article}
\def\be{\begin{eqnarray}}
\def\ee{\end{eqnarray}}
\def\ba{\begin{array}}
\def\ea{\end{array}}
\def\Z{{\cal Z}}
\def\C{{\cal C}}
\def\N{{\cal N}}
\def\E{{\cal E}}
\def\F{{\cal F}}
\begin{document}
\begin{center}
{\LARGE   { Heterotic string theory interrelations}}
\end{center}
\vskip 25mm
\begin{center}
{\bf \large {Oleg V. Kechkin}}
\end{center}
\begin{center}
DEPNI, Institute of Nuclear Physics\\
M.V. Lomonosov Moscow State University\\
119899 Moscow, Vorob'yovy Gory, Russia\\
e-mail:\, kechkin@depni.npi.msu.su
\end{center}
\vskip 25mm
\begin{abstract}
We establish a symmetry map which relates two low-energy heterotic string
theories with different numbers of the Abelian gauge fields compactified from
the diverse to three dimensions on a torus. We discuss two applications of the
established symmetry: a generation of the heterotic string theory solutions from
the stationary Einstein-Maxwell fields and one non-trivial submersion of the
heterotic string theory into the bosonic one.
\end{abstract}
\vskip 10mm
\begin{center}
\end{center}
%1111111111111111111111111111111111111111111111111111111111111
%\renewcommand{\theequation}{\thesection.\arabic{equation}}
\newpage
\section{Introduction}
Symmetries play a crucial role in the study of superstring theory \cite{Kir},
\cite{HullTow}, \cite{MahSch}, \cite{Sen4}, \cite{Sen3}. In this study the first
step is related to the search of hidden symmetries of the different field theory
limits of superstring theory. After that it becomes possible to make the second
step - to generalize the obtained results to the case of exact superstring
theory by exploring of the supersymmetry arguments \cite{Sensurv}, \cite{Kal},
\cite{Card}.

The heterotic string theory becomes a field theory of its massless modes at low
energies. The bosonic sector of this field theory leads to the symmetric space
model coupled to gravity after the toroidal compactification to three dimensions
\cite{MahSch}, \cite{Sen3}. This is the symmetric space model with the
null-curvature matrix parameterizing the coset $O(d+1, d+1+n)/O(d+1)\times
O(d+1+n)$, where $d$ is the number of compactified dimensions and $n$ is the
number of Abelian gauge fields in the multidimensional theory. In fact one has
a series of the low-energy heterotic string theory systems labeled by the
numbers $d$ and $n$, and all these field theories are of the interest in
framework of the basic exact heterotic string theory activity. Of course, the
special cases of $d=7, n=16$ (the consistent heterotic string theory) and $d=23,
n=0$ (the consistent bosonic string theory) are the most physically important.
However, some supergravities and the general superstring theory web of dualities
give a sufficient motivation for the study of whole effective set of the
$(d,n)$-labeled heterotic string theories.

In this paper we establish a symmetry map which relates two different
representatives of the $(d,n)$-series of the theories. This map is the on-shell
symmetry, so one can use it also as some new symmetry extension of the simpler
theory to the more complicated one (as a rule, in concrete work the simpler
theory is characterized by the smaller values of $d$ and $n$). Below we show
how one
can extend the solution space of the stationary Einstein-Maxwell theory \cite{EMT}
to the one of heterotic string theory with the arbitrary numbers $d$ and $n$.
This result opens a new and actually promising way for generation of the
heterotic string theory solutions (see \cite{Our} for different examples of
generation starting from the General Relativity).
Another interesting application of the
established symmetry map is related to some simple submersion of the heterotic
string theory into the bosonic one. This submersion seems surprising in
framework
of the conventional subsystem search for the following generation procedure.
Actually, in view of this submersion one generates the solutions of the theory
without Abelian fields from the solutions of the theory with non-trivial Abelian
gauge sector. Also it seems physically interesting that the critical case of the
heterotic string theory exactly transforms to the critical bosonic string theory one
in framework of the simplest new map realization.

%2222222222222222222222222222222222222222222222222222222222222222222
\section{Symmetric space model}

In \cite{ZF} it was shown that the low-energy heterotic string theory toroidally
compactified to three dimensions can be represented in terms of the single real
$(d+1)\times (d+1+n)$ matrix potential $\Z$ coupled to the effective
$3$-dimensional metric $h_{\mu\nu}$. The corresponding action reads:
\be\label{s1}
S_3&=&\int d^3x h^{1/2}\left ( -R_3+L_3\right ),
\nonumber\\
L_3&=&{\rm Tr}\left [ \nabla\Z\left (\Xi-\Z^T\Sigma\Z\right )^{-1}\nabla\Z^T
\left (\Sigma-\Z\Xi\Z^T\right )^{-1}\right ],
\ee
where $\Sigma$ and $\Xi$ are the $(d+1)\times (d+1)$ and $(d+1+n)\times (d+1+n)$
matrices respectively of the form ${\rm diag}(-1,-1;1,...,1)$ (see \cite{ZF} for
the details). It is important to note, that the three remaining coordinates
$x^{\mu}$ are Euclidean, the time dimension is also compactified together with
the `true' extra dimensions. Another important fact is that the
potential $\Z$ realizes the matrix representation of the lowest possible matrix
dimensionality: the symmetric space model $O(d+1, d+1+n)/O(d+1)\times O(d+1+n)$
has
exactly $(d+1)\times (d+1+n)$ degrees of freedom. Then, from Eq. (\ref{s1}) it
immediately follows that the transformation
\be\label{s2}
\Z\rightarrow\C_L\Z\C_R
\ee
is a symmetry if
\be\label{s3}
\C_L\Sigma\C_L=\Sigma, \quad \C_R\Xi\C_R=\Xi,
\ee
i.e., if $\C_L\in O(2,d-1)$ and $\C_R \in O(2,d-1+n)$. In \cite{HKCS} it was
shown that the group $O(2,d-1)\times O(2,d-1+n)$ forms the total subgroup of
charging symmetries for the theory under consideration. This subgroup preserves
the trivial values of the spatial asymptotics of the all three-dimensional
fields (i.e. the value $\Z=0$). These symmetries generalize the subgroup of the
Einstein-Maxwell charging symmetries \cite{EMT} to the heterotic string theory
case. It is important to note, that the reconstruction of the multidimensional
fields from the asymptotically flat three-dimensional solution can lead to NUT,
magnetic and other kind of Dirac string peculiarities. In all other senses such
multidimensional solution is also asymptotically trivial. The asymptotically
flat solutions of heterotic string theory play a crucial role in the black hole
physics \cite{Youm} and in the other physical applications of the superstring
theory.

The $\Z$-formulation is closely related to the null-curvature matrix approach
originally developed in \cite{Sen3}. Here the history is the following: in
\cite{HKMEP} it was constructed some new null-curvature matrix which is closely
related to the Ernst matrix potential approach. This approach establishes the
explicit analogy between the Einstein-Maxwell and heterotic string theories (see
the conventional Ernst potential representation of the stationary
Einstein-Maxwell theory in
\cite{Ernst}). In \cite{HKCS} the potential $\Z$ was initially introduced for
the simple linearization of action of the heterotic string theory charging
symmetries. After that in \cite{ZF} the $\Z$-formalism had been developed in
details as the new and powerful consistent approach. In this article we give,
may be, the most important application of this new approach to the physically
and mathematically interesting problems of the low-energy heterotic string
theory.

%33333333333333333333333333333333333333333333333333333333333333333333333333333

\section{New symmetry map}

The new symmetry map naturally arises in framework of one simple anzats
consideration. Let us start with
\be\label{m1}
\Z=\xi_{rl}L_lR_r^T,
\ee
where $l=1,...,\N_l,\,r=1,...,\N_r$ \,\, $\xi_{rl}=\xi_{rl}(x^{\mu})$ is the set
of $\N_l\times\N_r$ dynamical functions and $L_l, R_r$ are the constant columns
of the dimensions $(d+1)\times 1$ and $(d+1+n)\times 1$ respectively. The
question is when the dynamical equations
\be\label{m2}
&&\nabla^2\Z+2\nabla\Z\Xi\Z^T\left (\Sigma-\Z\Xi\Z^T\right )^{-1}\nabla\Z=0,
\nonumber\\
&&R_{3\,\mu\nu}={\rm Tr}\left [\Z_{,(\mu}\left (\Xi-\Z^T\Sigma\Z\right )^{-1}
\Z^T_{,\nu)}\left (\Sigma-\Z\Xi\Z^T\right )^{-1}\right],
\ee
which correspond to the action (\ref{s1}), are automatically satisfied in
framework of the structure (\ref{m1}). To answer on this question, one needs in
substitution of Eq. (\ref{m1}) to Eq. (\ref{m2}); this work can be elegantly
performed in terms of the operators
\be\label{m3}
\Pi_{l_1l_2}^{(L)}=L_{l_1}L_{l_2}^T\Sigma,\quad
\Pi_{r_1r_2}^{(R)}=R_{r_1}R_{r_2}^T\Xi
\ee
and the related constants
\be\label{m4}
\kappa_{l_1l_2}^{(L)}=L_{l_1}^T\Sigma L_{l_2},
\quad
\kappa_{r_1r_2}^{(R)}=R_{r_1}^T\Xi R_{r_2},
\ee
which are also related by the following multiplication table:
\be\label{m5}
\Pi_{l_1l_2}^{(L)}\Pi_{l_3l_4}^{(L)}=\kappa_{l_2l_3}^{(L)}\Pi_{l_1l_4}^{(L)},
\quad
\Pi_{r_1r_2}^{(R)}\Pi_{r_3r_4}^{(R)}=\kappa_{r_2r_3}^{(R)}\Pi_{r_1r_4}^{(R)}.
\ee
These relations characterize the operators (\ref{m3}) as projectors. In view of
this fact our approach can be naturally named as `projective'; this name will be
also supported by the following consideration and by the main result. Let us now
combine the quantities $\xi_{rl}, \kappa_{l_1l_2}^{(L)}$ and
$\kappa_{r_1r_2}^{(R)}$ into the matrices $\xi$, $\kappa^{(L)}$ and
$\kappa^{(R)}$ respectively. Then the straightforward calculation show that the
motion equations (\ref{m2}) become automatically satisfied if
\be\label{m6}
&&\nabla^2\xi+2\nabla\xi\kappa^{(R)}\xi^T\left ( [\kappa^{(L)}]^{-1}-
\xi\kappa^{(R)}\xi^T\right )^{-1}\nabla\xi=0,
\nonumber\\
&&R_{3\,\mu\nu}={\rm Tr}\left [\xi_{,(\mu}\left ( [\kappa^{(R)}]^{-1}-\xi^T
[\kappa^{(L)}]\xi\right )^{-1}
\xi^T_{,\nu)}\left ( [\kappa^{(L)}]^{-1}-\xi \kappa^{(R)}\xi^T\right )^{-1}
\right],
\ee
where it was supposed that the matrices $\kappa^{(L)}$ and $\kappa^{(R)}$ are
non-degenerated.

We state that the equations (\ref{m6}) are of the form (\ref{m2}) but written in
the `random' terms. Actually, let us denote the signature matrices for
$\kappa^{(L)}$ and $\kappa^{(R)}$ as $\tilde\Sigma$ and $\tilde\Xi$
respectively. Then from the corresponding algebraic theorem it follows an
existence of the matrices $\aleph^{(L)}$ and $\aleph^{(R)}$ such that
\be\label{m7}
\kappa^{(L)}=\aleph^{(L)\,T}\tilde\Sigma\aleph^{(L)},\quad
\kappa^{(R)}=\aleph^{(R)\,T}\tilde\Xi\aleph^{(R)}.
\ee
Let us also define the new dynamical field
\be\label{m8}
\tilde\Z=\aleph^{(L)}\xi\aleph^{(R)\,T}.
\ee
Then, as it is easy to verify, from Eq. (\ref{m6}) it follows that
\be\label{m9}
&&\nabla^2\tilde\Z+2\nabla\tilde\Z\tilde\Xi\tilde\Z^T\left (\tilde\Sigma-
\tilde\Z\tilde\Xi\tilde\Z^T\right )^{-1}\nabla\tilde\Z=0,
\nonumber\\
&&R_{3\,\mu\nu}={\rm Tr}\left [\tilde\Z_{,(\mu}\left (\tilde\Xi-\tilde\Z^T
\tilde\Sigma\tilde\Z\right )^{-1}
\tilde\Z^T_{,\nu)}\left (\tilde\Sigma-\tilde\Z\tilde\Xi\tilde\Z^T\right )^{-1}
\right],
\ee
i.e., the system which is related to (\ref{m2}) by the help of substitution
\be\label{m10}
\tilde\Z\leftrightarrow\Z,\quad \tilde\Sigma\leftrightarrow\Sigma,\quad
\tilde\Xi\leftrightarrow\Xi.
\ee
This system can be derived from the Lagrangian
\be\label{m11}
\tilde L_3={\rm Tr}\left [ \nabla\tilde\Z\left (\tilde\Xi-\tilde\Z^T\tilde
\Sigma\tilde\Z\right )^{-1}\tilde\Z^T
\left (\tilde\Sigma-\tilde\Z\tilde\Xi\tilde\Z^T\right )^{-1}\right ],
\ee
which has exactly the heterotic string theory form (\ref{s1}). The necessary
additional element of the correspondence (\ref{m10}) follows from Eqs.
(\ref{m1}), (\ref{m4}). To obtain it in the convenient form, let us define the
new set of constant columns
\be\label{m12}
L_{0l_1}=\left ( [\aleph^{(L)}]^{-1}\right )_{l_2l_1}L_{l_2}, \quad
R_{0r_1}=\left ( [\aleph^{(R)}]^{-1}\right )_{r_2r_1}R_{r_2}.
\ee
Then from Eq. (\ref{m4}) one obtains that
\be\label{m13}
L_{0l_1}^T\Sigma L_{0l_2}^T=\tilde\Sigma_{l_1l_2},\quad
R_{0r_1}^T\Xi R_{0r_2}^T=\tilde\Xi_{r_1r_2}.
\ee
Let us now combine the columns $L_{0l}$ and $R_{0r}$ to the matrices $L$ and
$R$ respectively in the natural way (then, for example, $L_{0l}$ is the $l$-th
column of $L$). Then Eq. (\ref{m13}) takes the form of
\be\label{m14}
L^T\Sigma L=\tilde\Sigma,\quad R^T\Xi R=\tilde\Xi,
\ee
whereas Eq. (\ref{m1}) reads:
\be\label{m15}
\Z=L\tilde\Z R^T.
\ee
The equations (\ref{s2}), (\ref{m11}), (\ref{m14}) and (\ref{m15}) define
the correspondence (\ref{m10}) completely. A simple algebraical analysis of Eq.
(\ref{m14}) shows that $\N_L\leq d+1$, $\N_R\leq d+1+n$ and the number of $-1$
in $\tilde\Sigma$ and $\tilde\Xi$ is not more than $2$. We would like to
interpret the $\tilde\Z$-theory as some other example of low-energy heterotic
string theory. To do this
let us completely define our anzats by the demanding that $\N_L\leq\N_R$ and
that both the matrices $\tilde\Sigma$ and $\tilde\Xi$ are of the form
${\rm diag}(-1,-1;1,...,1)$. Then the $\tilde\Z$-theory actually becomes the
heterotic
string theory with $\tilde d=\N_L-1$ toroidally compactified dimensions and
$\tilde n=\N_R-\N_L$ original Abelian gauge fields. It is easy to see that the
correspondence (\ref{m10}) becomes the general map which acts in the whole
$(d,n)$-labeled series of the low-energy heterotic string theory.

%4444444444444444444444444444444444444444444444444444444444444444444444444444444444

\section{Some applications}
Our first example is the $\tilde\Z$-theory with $\tilde n=2(\tilde d+1)$ and
the $\Z$-theory with $n=0, \, d=3\tilde d+2$. This special case includes the
critical heterotic and bosonic string theories ($\tilde d=7,\tilde n=16$ and
$d=23, n=0$ respectively). We state that this special case can be considered in
framework of the correspondence (\ref{m10}). Actually, let us take the matrices
$L$ and $R$ as
\be\label{e1}
L=\left (\ba{c}
1\cr
0
\ea\right ), \quad
R=1,
\ee
where the block unit matrix in $L$ has the dimension $(\tilde d+1)\times
(\tilde d+1)$. Then the relation
\be\label{e2}
\Z=\left (\ba{c}
\tilde\Z\cr
0
\ea\right )
\ee
is the realization of Eq. (\ref{m15}) in this situation. Note, that Eq.
(\ref{e2}) transform any solution of the heterotic string theory to the
corresponding solution of the bosonic string theory, including the critical
cases.

Our second example is the following. Let us put $\tilde d=1, \tilde n=2$
(this theory is interesting itself from the point of view of $D=N=4$
supergravity, see \cite{Kal} and references therein). In this case $\tilde\Z$ is
the $2\times 4$ matrix. Let us separate it to two $2\times 2$ blocks,
\be\label{e3}
\tilde\Z=(\tilde\Z_1,\tilde\Z_2),
\ee
and after that let us take these blocks in the following special form:
\be\label{e4}
\tilde\Z_a=\left (\ba{cc}
z_a^{'}&z_a^{''}\cr
-z_a^{''}&z_a^{'}
\ea\right ),
\ee
where $a=1,2$. In fact Eq. (\ref{e4}) gives the $2\times 2$ matrix
representation of the complex potentials
\be\label{e5}
z_a=z_a^{'}+iz_a^{''};
\ee
it is easy to prove that the anzats (\ref{e4}) is consistent, i.e., the
dynamical equations (\ref{m9}) do not impose any additional restrictions on the
potentials $z_a$. These equations become the equations on the complex
$1\times 2$ matrix potential $z$,
\be\label{e6}
z=(z_1, z_2),
\ee
and can be derived from the action (\ref{s1}) with the Lagrangian
\be\label{e7}
L_{EM}=2\frac{\nabla z (\sigma_3-z^+z)^{-1}\nabla z^+}{1-z\sigma_3z^+},
\ee
where $\sigma_3$ is one of the Pauli matrices. We state that this effective
theory coincides with the stationary Einstein-Maxwell one. To prove this
statement, let us introduce the potentials $\E$ and $\F$, where
\be\label{e8}
\E=\frac{1-z_1}{1+z_1}, \quad \F=\frac{\sqrt 2z_2}{1+z_1}.
\ee
Then for the Lagrangian (\ref{e6}) one has:
\be\label{e9}
L_{EM}=\frac{1}{2f^2}\left | \nabla\E-\bar\F\nabla\F\right |^2-
\frac{1}{f}\left |\nabla\F\right |^2,
\ee
where $f=1/2(\E+\bar{\E}-|\F|^2)$. It is clear that (\ref{e9}) is the
conventional action for the stationary Einstein-Maxwell theory written in terms
of the Ernst potentials $\E$ and $\F$ (see \cite{Ernst}). Thus, Eqs.
(\ref{e3})-(\ref{e7}) define the $\tilde\Z$-system which is equivalent to the
stationary Einstein-Maxwell theory. It can be used for generation of the
heterotic string theory solutions from the stationary Einstein-Maxwell ones
by the help of the above established symmetry map. Of course, all these results
can be easily modified to the `cosmological' axisymmetric case when the
effective three-dimensional coordinate space contains the time dimension.

%GGGGGGGGGGGGGGGGGGGGGGGGGGGGGGGGGGGGGGGGGGGGGGGGGGGGGGGGGGGGGGGGGGGGG

\section*{Acknowledgments}
This work was supported by RFBR grant ${\rm N^{0}}
\,\, 00\,02\,17135$.

%444444444444444444444444444444444444444444444444444444444444444444444

\section{Conclusion}

A simplicity which characterizes the process of results obtaining is related to
remarkable properties of the $\Z$-representation of the low-energy heterotic
string theory compactified to three dimensions on a torus. A very similar
formalism for the stationary Einstein-Maxwell theory one can find in \cite{Maz},
where the equivalents of our potentials $z_a$ had been introduced. It is
important
to note that the $\Z$-approach is especially useful in framework of the
asymptotically flat solutions of the theory. In fact our symmetry map
(\ref{m15}) coincides with the general charging symmetry transformation
\cite{HKCS} in the case when $\Z$ and $\tilde\Z$ are of the same matrix
dimensionality (see Eq. (\ref{s2})). Moreover, in the general case of the
different dimensionalities of the potentials $\Z$ and $\tilde\Z$ our symmetry
map possessing the charging symmetry invariance property itself. This means the
following: an attempt to generalize the result given by Eq. (\ref{m15}) by the
help of the transformation (\ref{s2}) applied to the $\Z$ and $\tilde\Z$
potentials leads to the non-important reparameterization of the constant
matrices
$L$ and $R$. Actually, it is easy to see that this reparameterization preserves
the restrictions (\ref{m14}), so it play the zero role if one takes the general
solution of the algebraic equations (\ref{m14}).

At the end of this paper let us briefly discuss some natural perspectives.
First of all, it is possible to 'translate' the material given in the previous
section from the language of three-dimensional potentials to the form of
physical field components. All the necessary information for this 'translation'
can be found in \cite{HKCS} and \cite{ZF}; in this short paper we will not
give all
important but technically complicated details. Second, it will be interesting
to calculate the concrete heterotic string theory solutions from the known
Einstein-Maxwell theory ones using our new symmetry map. For example, it is
possible
to construct the charging symmetry complete extension of the Kerr-Newman
solution. The corresponding work is now in progress.

%LLLLLLLLLLLLLLLLLLLLLLLLLLLLLLLLLLLLLLLLLLLLLLLLLLLLLLLLLLLLLL


\begin{thebibliography}{23}

\bibitem{Kir}
E. Kiritsis,
``Introduction to superstring theory'',
Leuven Univ. Pr. (1998).

\bibitem{HullTow}
C.M. Hull, P.K. Townsend,
Nucl. Phys. {\bf B438} (1995) 109.

\bibitem{MahSch}
J. Maharana, J.H. Schwarz,
Nucl. Phys. {\bf B390} (1993) 3.

\bibitem{Sen4}
A. Sen,
Int. J. Mod. Phys. {\bf A9} (1994) 3707.

\bibitem{Sen3}
A. Sen,
Nucl. Phys. {\bf B434} (1995) 179.

\bibitem{Sensurv}
A. Sen,
`An introduction to nonperturbative string theory',
in `Cambridge 1997, Duality and supersymmetric theories' 297.

\bibitem{Kal}
E. Bergshoeff, d. Kastor, T. Ortin,
Nucl. Phys. {\bf B478} (1996) 156.

\bibitem{Card}
I. Bakas, M. Bourdeau, G. L. Cardoso,
Nucl. Phys. {\bf B510} (1998) 103.

\bibitem{EMT}
D. Kramer, H. Stephani, M. MacCallum, E. Herlt,
``Exact solutions of the Einstein field equations'',
Deutcher Verlag der Wissenschaften, Berlin, (1980).

\bibitem{Our}
O.V. Kechkin, M.V. Yurova,
Gen Rel. Grav. {\bf 29} (1997) 1283;
D.V. Galtsov, O.V. Kechkin,
Phys. Rev. {\bf D50} (1994) 7394;
D.V. Galtsov, O.V. Kechkin,
Phys. Lett. {\bf B361} (1995) 52.

\bibitem{ZF}
O.V. Kechkin,
``New matrix formalism for heterotic string theory
on a torus'' gr-qc/0012090.

\bibitem{HKCS}
A. Herrera-Aguilar, O.V. Kechkin,
Phys. Rev. {\bf D59} (1999) 124006.

\bibitem{Youm}
D. Youm,
Phys. Rept. {\bf 316} (1999) 1.

\bibitem{HKMEP}
A. Herrera-Aguilar, O.V. Kechkin,
Int. J. Mod. Phys. {\bf A13} (1998) 393.

\bibitem{Ernst}
F. J. Ernst,
Phys. Rev. {\bf 167} 5 (1968) 1175.


\bibitem{Maz}
P.O. Mazur,
Acta Phys. Pol. {\bf B14} (1983) 219.

\end{thebibliography}
\end{document}